\documentclass[11pt]{article}
\usepackage{mathrsfs}
\usepackage[all,pdf]{xy}
\usepackage{amsfonts}
\usepackage{lscape}
\usepackage{amssymb}
\usepackage{bbm}
\usepackage{color}
\usepackage{amsfonts,amssymb, mathrsfs, amsmath,   amssymb,  theorem,  float}
\usepackage{graphicx}  %% note spelling
\newtheorem{theorem}{Theorem}[section]

\newtheorem{cor}[theorem]{Corollary}
\newtheorem{lemma}[theorem]{Lemma}
\newtheorem{alg}[theorem]{Algorithm}
\newtheorem{remark}[theorem]{Remark}

\def\qed{\hfil {\vrule height5pt width2pt depth2pt}}

\def\qed{\hfil {\vrule height5pt width2pt depth2pt}}
\def\bref#1{(\ref{#1})}

\def\qed{\hfil {\vrule height5pt width2pt depth2pt}}

\def\C{\mathcal{C}}

\def\proof{{\noindent\em Proof.\,\,}}

\def\bref#1{(\ref{#1})}

\def\N{{\mathbb N}}
\def\Q{{\mathbb Q}}

\def\C{{\mathbb C}}

\def\bref#1{(\ref{#1})}

\def\Z{{\mathbb Z}}
\def\C{{\mathbb C}}

\def\+{ \oplus}
\def\-{\ominus}
\def\*{\otimes}

\def\deg{\hbox{\rm{deg}}}

\begin{document}
% \clubpenalty=10000
% \widowpenalty = 10000

%\thispagesyle{empty}

%\thanks{\quad Partially
%       supported by a National Key Basic Research Project of China %(2011CB302400) and  by a grant from NSFC (60821002).}}

\title{Sparse Polynomial Interpolation  with\\ Finitely Many Values for the Coefficients\thanks{Partially supported by a grant from NSFC No.11688101.}}
\author{Qiaolong Huang and Xiao-Shan Gao \\
 KLMM, UCAS,  Academy of Mathematics and Systems Science\\
 Chinese Academy of Sciences, Beijing 100190, China}
\date{}

\maketitle

\begin{abstract}
\noindent
In this paper, we give new sparse interpolation algorithms for
black box   polynomial $f$
whose coefficients  are from a finite set.
In the univariate case, we recover $f$ from one evaluation
$f(\beta)$ for a sufficiently large number $\beta$.
In the multivariate case, we introduce the modified Kronecker substitution
to reduce the interpolation of a multivariate polynomial to
that of the univariate case. Both algorithms have polynomial bit-size complexity.

\end{abstract}

\section{Introduction}
The interpolation for a sparse multivariate polynomial
 $f(x_1,\ldots,x_n)$
given as a black box is a basic computational problem.
Interpolation algorithms were given when we know an
upper bound for the terms of $f$ \cite{1}
and upper bounds for the terms and the degrees of $f$ \cite{71}.
These algorithms were significantly improved and these works
can be found in the references of \cite{72}.

In this paper, we consider the sparse interpolation for $f$
whose coefficients are taken from a known finite set.
For example, $f$ could be in $\Z[x_1,\ldots,x_n]$ with an upper bound on the absolute values of coefficients of $f$, or $f$ is in $\Q[x_1,\ldots,x_n]$ with upper bounds both on the absolute values of coefficients and their denominators.

This kind of interpolation is motivated by the following applications.
The interpolation of sparse rational functions leads to
interpolation of sparse polynomials whose coefficients have bounded denominators \cite[p.6]{812}.
In \cite{8123}, a new method is introduced to reduce the interpolation of a multivariate polynomial $f$ into the interpolation of univariate polynomials, where we need to obtain the terms of $f$ from a larger set of terms and the method given in this paper is needed to solve this problem.

In the univariate case, we show that if $\beta$ is larger than a given
bound depending on the coefficients of $f$,
then  $f$ can be recovered from $f(\beta)$.
Based on this idea, we give a sparse interpolation algorithm
for univariate polynomials with rational numbers as coefficients,
whose bit complexity is
$\mathcal{O}((td\log H(\log C+\log H))$ or $\widetilde{\mathcal{O}}(td)$,
where $t$ is the number of terms of $f$, $d$ is the degree of $f$,
$C$ and $H$ are upper bounds for the coefficients and
the denominators of the coefficients of $f$.
%
%One advantage of the algorithm is that only $C$ and $H$ are required in the input.
It seems that the algorithm has the optimal bit complexity $\widetilde{\mathcal{O}}(td)$ in all known deterministic and exact interpolation algorithms for black box univariate
polynomials as discussed in Remark \ref{rem-1}.

In the multivariate case, we show that by choosing a good prime,
the interpolation of a multivariate polynomial can be reduced to
that of the univariate case in polynomial-time. As a consequence,
a new sparse interpolation algorithm for  multivariate polynomials
is given, which has  polynomial bit-size complexity.  We also give its probabilistic version.

There exist many methods for reducing the interpolation of a multivariate  polynomial into that of univariate polynomials, like the classical Kronecker substitution, randomize Kronecker substitutions\cite{7}, Zipple's algorithm\cite{71}, Klivans-Spielman's algorithm\cite{5}, Garg-Schost's algorithm \cite{21}, and Giesbrecht-Roche's algorithm\cite{6}.
Using the original Kronecker substitution \cite{2}, interpolation for multivariate polynomials  can be easily reduced to the univariate case.
The main problem with this approach is that the highest degree of the univariate polynomial  and the height of the data in the algorithm are exponential.
In this paper, we give the following modified Kronecker substitution
$$x_i=x^{\mathbf{mod}((D+1)^{i-1},p)},i=1,2,\dots,n$$
to reduce multivariate interpolations to univariate interpolations.
Our approach simplifies and builds on previous work by Garg-Schost\cite{21}, Giesbrecht-Roche\cite{6}, and Klivans-Spielman\cite{5}. The first two are for straight-line programs. Our interpolation algorithm works for the more general setting of black box sampling.

The rest of this paper is organized as follows. In Section 2, we give interpolation algorithms about univariate polynomials.
In Section 3, we give interpolation algorithms about multivariate polynomials.
In Section 4, experimental results are presented.

\section{Univariate polynomial interpolation}

\subsection{Sparse interpolation with finitely many coefficients}

In this section, we always assume
\begin{equation}\label{eq-f}
f(x)=c_1x^{d_1}+c_2x^{d_2}+\dots+c_tx^{d_t}\end{equation}
where $d_1,d_2,\dots,d_t\in \N,d_1<d_2<\cdots<d_t$, and $c_1,c_2,\cdots,c_t\in A$, where $A\subset \mathbb{C}$ is a finite set.
Introduce the following notations
\begin{equation}\label{eq-e}
C:=\max_{a\in A}(|a|),\quad \varepsilon:=\min (\varepsilon_1,\varepsilon_2)
\end{equation}
where
$\varepsilon_1 :=\min_{a,b\in A,a\neq b}|a-b|$ and $\varepsilon_2 :=\min_{a\in A,a\neq 0} |a|$.

\begin{theorem}\label{the-5}
If $\beta\geq \frac{2C}{\varepsilon}+1$, then $f(x)$ can be uniquely determined by $f(\beta)$.
\end{theorem}

\proof
Firstly, for $\forall k=1,2,\cdots$, we have
\begin{align}
\beta\geq \frac{2C}{\varepsilon}+1\Longrightarrow{} & \beta-1\geq \frac{2C}{\varepsilon}\notag \\
\Longrightarrow{} & \beta-1>\frac{2C}{\varepsilon}\frac{\beta^k-1}{\beta^k}\notag\\
\Longrightarrow{} & \varepsilon \beta^k>2C\frac{\beta^k-1}{\beta-1}\notag\\
\Longrightarrow{} & \varepsilon \beta^k>2C(\beta^{k-1}+\beta^{k-2}+\cdots+\beta+1)\notag
\end{align}
From \bref{eq-f}, we have $f(\beta)=c_1\beta^{d_1}+c_2\beta^{d_2}+\cdots+c_t\beta^{d_t}$.
Assume that there is another form $f(\beta)=a_1\beta^{k_1}+a_2\beta^{k_2}+\dots+a_s\beta^{k_s}$, where $a_1,a_1,\dots,a_s \in A$
and $k_1<k_2<\cdots<k_s$. It suffices to show that $c_t\beta^{d_t}=a_s\beta^{k_s}$. The rest can be proved by induction.
First assume that $d_t\neq k_s$. Without loss of generality, let $d_t>k_s$. Then we have
\begin{align}
0=&|(c_1\beta^{d_1}+c_2\beta^{d_2}+\cdots+c_t\beta^{d_t})-(a_1\beta^{k_1}+a_2\beta^{k_2}+\cdots+a_s\beta^{k_s})|\notag\\
\geq{} & |c_t| \beta^{d_t}-C(\beta^{d_t-1}+\cdots+\beta+1)-C(\beta^{k_s}+\cdots+\beta+1)\notag\\
\geq{} &|c_t| \beta^{d_t}-2C(\beta^{d_t-1}+\cdots+\beta+1)\notag\\
>{}  &|c_t| \beta^{d_t}-\varepsilon \beta^{d_t}\geq0\notag
 \end{align}
It is a contradiction, so $d_t=k_s$.
Assume $c_t\neq a_s$, then
\begin{align}
 0=&|(c_1\beta^{d_1}+c_2\beta^{d_2}+\cdots+c_t\beta^{d_t})-(a_1\beta^{k_1}+a_2\beta^{k_2}+\cdots+a_s\beta^{k_s})\notag\\
\geq{} &|c_t-a_s| \beta^{d_t}-2C(\beta^{d_t-1}+\cdots+\beta+1)\notag\\
>{} &|c_t-a_s| \beta^{d_t}-\varepsilon \beta^{d_t}\geq0\notag
\end{align}
It is a contradiction, so $c_t=a_s$.
The theorem has been proved. \qed

\subsection{The sparse interpolation algorithm}
%\subsection{The sparse interpolation algorithm}
The idea of the algorithm is first to obtain the maximum term $m$ of $f$, then  subtract $m(\beta)$ from $f(\beta)$ and repeat the procedure until $f(\beta)$ becomes $0$.

We first show how to compute the leading degree $d_t$.
\begin{lemma}\label{the-1}
If $\beta\geq \frac{2C}{\varepsilon}+1$, then
%\begin{equation}
$|\frac{f(\beta)}{\beta^k}|=\begin{cases}
>\frac{\varepsilon}{2},&\text{if }  k\leq d_t\\
<\frac{\varepsilon}{2},&\text{if }  k> d_t
\end{cases}
$
%\end{equation}
\end{lemma}
\proof
From
%\begin{align}
$|f(\beta)|=|c_1\beta^{d_1}+c_2\beta^{d_2}+\cdots+c_t\beta^{d_t}|%\notag\\
\leq{}  C(\beta^{d_t}+\cdots+\beta+1)%\notag\\
={} C(\frac{\beta^{d_t+1}-1}{\beta-1})$
%\end{align}
and
%\begin{align}
$|f(\beta)|={} |c_1\beta^{d_1}+c_2\beta^{d_2}+\cdots+c_t\beta^{d_t}|
\geq{}  |c_t|\beta^{d_t}-C(\beta^{d_t-1}+\cdots+\beta+1)
={} |c_t|\beta^{d_t}-C\frac{\beta^{d_t}-1}{\beta-1},$
%\end{align}
we have
\begin{equation*}
|c_t|\beta^{d_t}-C\frac{\beta^{d_t}-1}{\beta-1}\leq |f(\beta)| \leq C (\frac{\beta^{d_t+1}-1}{\beta-1}).
\end{equation*}
When $k\leq d_t$,
$|\frac{f(\beta)}{\beta^k}|\geq{} |c_t|\beta^{d_t-k}-\frac{C}{\beta-1}(\beta^{d_t-k}-\frac{1}{\beta^k})
\geq{} \varepsilon \beta^{d_t-k}-\frac{\varepsilon}{2}(\beta^{d_t-k}-\frac{1}{\beta^k})
\geq{} \frac{\varepsilon}{2}\beta^{d_t-k}+\frac{\varepsilon}{2}\frac{1}{\beta^k}>\frac{\varepsilon}{2}$.
When $k>d_t$,
$|\frac{f(\beta)}{\beta^k}|\leq{}  \frac{C}{\beta-1}(\beta^{d_t+1-k}-\frac{1}{\beta^k})
\leq{} \frac{\varepsilon}{2}(\beta^{d_t+1-k}-\frac{1}{\beta^k})
\leq{} \frac{\varepsilon}{2}\beta^{d_t+1-k}-\frac{\varepsilon}{2}\frac{1}{\beta^k}<\frac{\varepsilon}{2}$.
%
%When $k\leq d_t$,
%\begin{align}
%|\frac{f(p)}{p^k}|\geq{} & |c_t|p^{d_t-k}-\frac{C}{p-1}(p^{d_t-k}-\frac{1}{p^k})\notag\\
%\geq{} & \varepsilon p^{d_t-k}-\frac{\varepsilon}{2}(p^{d_t-k}-\frac{1}{p^k})\notag\\
%\geq{} & \frac{\varepsilon}{2}p^{d_t-k}+\frac{\varepsilon}{2}\frac{1}{p^k}>\frac{\varepsilon}{2}\notag
%\end{align}
%%
%When $k>d_t$,
%\begin{align}
%|\frac{f(p)}{p^k}|\leq{} & \frac{C}{p-1}(p^{d_t+1-k}-\frac{1}{p^k})\notag\\
%\leq{} &\frac{\varepsilon}{2}(p^{d_t+1-k}-\frac{1}{p^k})\notag\\
%\leq{} &\frac{\varepsilon}{2}p^{d_t+1-k}-\frac{\varepsilon}{2}\frac{1}{p^k}<\frac{\varepsilon}{2}\notag
%\end{align}
\qed

If we can use logarithm operation, we can change the above lemma into the following form.

\begin{lemma}\label{lm-1}
If $\beta\geq \frac{2C}{\varepsilon}+1$, then
$d_t=\lfloor\log_\beta \frac{2|f(\beta)|}{\varepsilon}\rfloor$.
\end{lemma}
\proof
By lemma \ref{the-1}, we know $\frac{|f(\beta)|}{\beta^{d_t}}>\frac{\varepsilon}{2}$ and $\frac{|f(\beta)|}{\beta^{d_t+1}}<\frac{\varepsilon}{2}$. Then we have $\log_\beta \frac{|f(\beta)|}{\beta^{d_t}}>\log_\beta \frac{\varepsilon}{2}$ and $\log_\beta\frac{|f(\beta)|}{\beta^{d_t+1}}<\log_\beta\frac{\varepsilon}{2}$, this can be reduced into $\log_\beta \frac{2|f(\beta)|}{\varepsilon}-1<d_t<\log_\beta \frac{2|f(\beta)|}{\varepsilon}$. As $d_t$ is an integer, then we have $d_t=\lfloor\log_\beta \frac{2|f(\beta)|}{\varepsilon}\rfloor$.\qed

Based on Lemma \ref{lm-1}, we have the following algorithm which will be used in several places.
\begin{alg}[UDeg]\label{alg-2}
\end{alg}

{\noindent\bf Input:} $f(\beta),\varepsilon,$ where $\beta\geq \frac{2C}{\varepsilon}+1$.

{\noindent\bf Output:} the degree  of $f(x)$.
\begin{description}
%\item[Step 1:] set $u:=\frac{2|f(p)|}{\varepsilon}$;
\item[Step 1:] return $\lfloor \log_\beta (\frac{2|f(\beta)|}{\varepsilon})\rfloor$.
\end{description}

\begin{remark}\label{rem-1}
If we cannot use logarithm operation, then it is easy to show that we need $\mathcal{O}(\log^2 D)$ arithmetic operations to obtain the degree based on Lemma \ref{the-1}.
In the following section, we will regard logarithm as a basic step.
\end{remark}

Now we will show how to compute the leading coefficient $c_t$.
\begin{lemma}\label{the-2}
If $\beta\geq \frac{2C}{\varepsilon}+1$, then $c_t$ is the only element in $A$ that satisfies $|\frac{f(\beta)}{\beta^{d_t}}-c_t|<\frac{\varepsilon}{2}$.

\end{lemma}

\proof
First we show that $c_t$ satisfies $|\frac{f(\beta)}{\beta^{d_t}}-c_t|<\frac{\varepsilon}{2}$.
We rewrite $f(\beta)$ as $f(\beta)=c_t\beta^{d_t}+g(\beta)$, where $g(x):=c_{t-1}x^{d_{t-1}}+c_{t-2}x^{d_{t-2}}+\cdots+c_1x^{d_1}$. So $\frac{f(\beta)}{\beta^{d_t}}=c_t+\frac{g(\beta)}{\beta^{d_t}}$. As $\deg(g)<d_t$, by Lemma \ref{the-1}, we have $|\frac{g(\beta)}{\beta^{d_t}}|<\frac{\varepsilon}{2}$. So $|\frac{f(\beta)}{\beta^{d_t}}-c_t|<\frac{\varepsilon}{2}$.

Assume there is another $c\in A$ also have $|\frac{f(\beta)}{\beta^{d_t}}-c|<\frac{\varepsilon}{2}$, then $|c_t-c|\leq |\frac{f(\beta)}{\beta^{d_t}}-c|+|\frac{f(\beta)}{\beta^{d_t}}-c_t|<\varepsilon$. This is only happen when $c_t=c$, so we prove the uniqueness.\qed

Based on Lemma \ref{the-2}, we give the algorithm to obtain the leading coefficient.
%The details of obtain the coefficient we will give later for different cases.
%
\begin{alg}[ULCoef]\label{alg-uvlc}
\end{alg}

{\noindent\bf Input:} $f(\beta),\beta,\varepsilon, d_t$

{\noindent\bf Output:} the leading coefficient of $f(x)$

\begin{description}
\item[Step 1:]
Find the element $c$ in $A$ such that $|\frac{f(\beta)}{\beta^{d_t}}-c|<\frac{\varepsilon}{2}
$.

\item[Step 2:] Return $c$.

\end{description}

Now we can give the complete algorithm.
\begin{alg}[UPolySI]\label{alg-1}
\end{alg}

{\noindent\bf Input:} A black box univariate polynomial $f(x)$, whose coefficients
are in $A$.

{\noindent\bf Output:} The exact form of $f(x)$.

\begin{description}
\item[Step 1:]
Find the bounds $C$ and $\varepsilon$ of $A$, as defined in \bref{eq-e}.

\item[Step 2:]
Let $\beta:=\frac{2C}{\varepsilon}+1$.

\item[Step 3:]
Let $g:=0,u:=f(\beta)$.

\item[Step 4:]

$\mathbf{while}$ $u\neq 0$ $\mathbf{do}$

 $d:=${\bf UDeg}$(u,\varepsilon,\beta)$;

 $c:=${\bf ULCoef}$(u,\beta,\varepsilon,d)$;

  $u:=u-c\beta^d$;

  $g:=g+cx^d$;

  $\mathbf{end\ do}$.

\item[Step 5:]
Return $g$.
\end{description}

%Now we analyse the complexity of the algorithm.
Note that, the complexity of Algorithm \ref{alg-uvlc} depends on $A$, which is denoted by $O_A$. Note that $O_A\le |A|$.
We have the following theorem.

\begin{theorem}
The arithmetic complexity of the Algorithm \ref{alg-1} is $\mathcal{O}(tO_A)\le \mathcal{O}(t|A|)$,
where $t$ is the number of terms in $f$.
\end{theorem}
\proof
Since finding the maximum degree needs one operation and finding the coefficient of the maximum term needs $O_A$ operations, and finding the maximum term needs $\mathcal{O}(O_A)$ operations. We prove the theorem.\qed

%As a simplest case, if the coefficients are taken from a bounded
%set of integers, we have the following results.
%\begin{theorem}\label{th-int1}
%If the coefficients of $f(x)$ are integers in $[\C,C]$, then
%Algorithm \ref{alg-1} can compute $f(x)$ with bit complexity $O()$.
%\end{theorem}
%
%In this case, $A=\{-C,-(C-1),\dots,-1,1,\dots,C-1,C\},\varepsilon=1$, so we can choose $p=2C+1$.
%%
%It is easy to see that find the coefficients is trivial, we design the algorithm in details as follows.
%
%
%
%{\noindent\bf Input:} $\frac{f(p)}{p^{d_t}}$
%
%{\noindent\bf Output:} the coefficient of $x^{d_t}$ in $f(x)$
%
%
%\begin{description}
%\item[Step 1:]let $u:=\frac{f(p)}{p^{d_t}}$.
%
%\item[Step 2:] Return $\lceil u- \frac12\rceil$.
%
%
%\end{description}
%
%\begin{alg}[CoefInt]
%\end{alg}
%
%{\noindent\bf Input:} $\frac{f(p)}{p^{d_t}}$
%
%{\noindent\bf Output:} the coefficient of $x^{d_t}$ in $f(x)$
%
%
%\begin{description}
%\item[Step 1:]let $u:=\frac{f(p)}{p^{d_t}}$.
%
%\item[Step 2:] Return $\lceil u- \frac12\rceil$.
%
%\end{description}

\subsection{The rational number coefficients case}

In this section, we assume that the coefficients of $f(x)$ are rational numbers in \begin{equation}\label{eq-rat}
A=\{\frac b a \,|\, 0<a\leq H,|\frac{b}{a}|\leq C,a,b\in \Z\}
\end{equation}
and we have $\varepsilon=\frac{1}{H(H-1)}$.
Notice that in Algorithm \ref{alg-1}, only Algorithm \ref{alg-uvlc} (${\bf ULCoef}$) needs refinement.
We first consider the following general problem about rational numbers.

\begin{lemma}\label{lm-uvrc2}
Let $0<r_1<r_2$ be rational numbers. Then we can find the smallest $d>0$
such that a rational number with denominator $d$ is in $(r_1,r_2)$
with computational complexity $\mathcal{O}(\log(r_2-r_1))$.
\end{lemma}
\proof
%We now give a method to find the denominator.
%Without loss of generality, we will assume that $q_1>0$.
%
We consider three cases.

$1$. If one of the $r_1$ and $r_2$ is an integer and the other one is not, then the smallest positive integer $d$ such that $(r_2-r_1)d>1$ is the smallest denominator, and $d=\lceil\frac{1}{r_2-r_1}\rceil$.

$2$. Both of $r_1,r_2$ are integers. If $r_2-r_1>1$, then $1$ is the smallest denominator. If $r_2-r_1=1$, then $2$ is the smallest denominator.

$3$. Both of $r_1,r_2$ are not integers. This is the most complicated case.

First, we check if there exists an integer in $(r_1,r_2)$.
If $\lceil r_1\rceil<r_2$,  then $\lceil r_1\rceil$ is in the interval which has the smallest denominator $1$.

Now we consider the case that $(r_1,r_2)$ does not contain an integer. Assume $r_1<\frac{d_1}{d}<r_2$, where $d>1$ is the smallest denominator.
Denote $w:=$trunc$(r_1)$, $\epsilon_1:=r_1-w$,$\epsilon_2:=r_2-w$. Then $\epsilon_1<\epsilon_2<1$ and $d$ is the smallest positive integer such that $(dr_1,dr_2)$ contains an integer.
Since  $dr_1=d(w+\epsilon_1),dr_2=d(w+\epsilon_2)$,  $d$ is the smallest positive integer such that interval $(d\epsilon_1,d\epsilon_2)$ contains an integer. Since  $dr_1=d(w+\epsilon_1),dr_2=d(w+\epsilon_2)$,  $d$ is the smallest positive integer such that interval $(d\epsilon_1,d\epsilon_2)$ contains an integer. We still denote it $d_1$. Then $d\epsilon_1<d_1< d\epsilon_2$, so $\frac{d_1}{\epsilon_2}< d<\frac{d_1}{\epsilon_1}$, and we can see that $d_1$ is the the smallest integer such that $(\frac{d_1}{\epsilon_2},\frac{d_1}{\epsilon_1})$ contains an integer.
Suppose we know how to compute the number $d_1$. Then $d=\lceil\frac{d_1}{\epsilon_2}\rceil$ when $\frac{d_1}{\epsilon_2}$ is not an integer, and $d=\frac{d_1}{\epsilon_2}+1$ when $\frac{d_1}{\epsilon_2}$ is an integer.

Note that $d_1$ is the smallest denominator such that some rational number
$\frac{d}{d_1}$ is in $(\frac{1}{\epsilon_2},\frac{1}{\epsilon_1})$.
To find $d_1$, we need to repeat the above procedure to $(\frac{1}{\epsilon_2},\frac{1}{\epsilon_1})$ and obtain a sequence of
intervals $(r_1,r_2)\rightarrow (\frac{1}{\epsilon_2},\frac{1}{\epsilon_1})\rightarrow \cdots$. The denominators of end points of the intervals becomes smaller after each repetition. So the algorithm will terminates.

Now we prove that the number of operations of the procedure
is $\mathcal{O}(\log(r_2-r_1))$. First, we know the length of the interval $(r_1,r_2)$ is $r_2-r_1$. Now we prove that every time we run one or two recursive steps, the length of the new interval will be $2$ times bigger.
Let $(\frac{b_1}{a_1},\frac{b_2}{a_2})$ be the first interval.
If it contains an integer, then we finish the algorithm. We assume that case does not happen,
so we can assume $|\frac{b_1}{a_1}|\leq1,|\frac{b_2}{a_2}|\leq1$. Then the second interval is $(\frac{a_2}{b_2},\frac{a_1}{b_1})$.
Now the new interval length is $\frac{a_1}{b_1}-\frac{a_2}{b_2}$.
If $\frac{b_1}{a_1}\leq\frac{1}{2}$, then we have
$\frac{\frac{a_1}{b_1}-\frac{a_2}{b_2}}{\frac{b_2}{a_2}-\frac{b_1}{a_1}}=\frac{\frac{a_1b_2-a_2b_1}{b_1b_2}}{\frac{a_1b_2-a_2b_1}{a_1a_2}}=\frac{a_1a_2}{b_1b_2}\geq2$.

If $\frac{b_1}{a_1}>\frac{1}{2}$, then we let $a_1=b_1+c_1,a_2=b_2+c_2$ and the third interval is $(\frac{b_1}{c_1},\frac{b_2}{c_2})$.

Then we have
$\frac{\frac{b_2}{c_2}-\frac{b_1}{c_1}}{\frac{b_2}{a_2}-\frac{b_1}{a_1}}=\frac{\frac{c_1b_2-c_2b_1}{c_1c_2}}{\frac{a_1b_2-a_2b_1}{a_1a_2}}=\frac{a_1a_2}{c_1c_2}>2$.
In this case, if we have an interval whose length is bigger than $1$, then the recursion will terminate. So if $(r_2-r_1)2^k\geq 1$, then $2k$ is the upper bound of the number of recursions. So the complexity is $\mathcal{O}(\log (r_2-r_1))$.
We proved the lemma.\qed

Based on Lemma \ref{lm-uvrc2}, we present a recursive algorithm to compute the rational number in an interval $(r_1,r_2)$ with the smallest denominator.
\begin{alg}[MiniDenom]\label{alg-mindn}
\end{alg}

{\noindent\bf Input:} $r_1,r_2$ are positive rational numbers.

{\noindent\bf Output:} the minimum denominator of rational numbers in $(r_1,r_2)$

\begin{description}
\item[Step 1:]$\mathbf{if}$ one of $r_1,r_2$ is an integer and the other one is not an integer $\mathbf{then}$ return $\lceil \frac{1}{r_2-r_1}\rceil$.

\item[Step 2:]$\mathbf{if}$ both of $r_1$ and $r_2$ are integers and $r_2-r_1>1$ $\mathbf{then}$ return $1$.

$\mathbf{if}$ both of $r_1$ and $r_2$ are integers and $r_2-r_1=1$ $\mathbf{then}$ return $2$.

\item[Step 3:]$\mathbf{if}$ $\lceil r_1\rceil<r_2$, $\mathbf{then}$ return 1.

\item[Step 4:]let $w:=$trunc$(r_1)$, $\epsilon_1:=r_1-w$,$\epsilon_2:=r_2-w$;

$d_1:=\mathbf{MiniDenom}(\frac{1}{\epsilon_2},\frac{1}{\epsilon_1})$;

$\mathbf{if}$ $\frac{d_1}{\epsilon_2}$ is a integer $\mathbf{then}$ return $\frac{d_1}{\epsilon_2}+1$
$\mathbf{else}$ return  $\lceil\frac{d_1}{{\epsilon_2}}\rceil$.

\end{description}

We now show how to compute the leading coefficient of $f(x)$.
\begin{lemma}\label{the-3}
Suppose $c_t=\frac{b}{a}$, where $\gcd(a,b)=1,a>0$, and $I_i=(\frac{f(\beta)}{\beta^{d_t}} i-\frac{\varepsilon}{2} i,\frac{f(\beta)}{\beta^{d_t}} i+\frac{\varepsilon}{2} i),i=1,2,\dots,H$.
Then $I_a\cap\Z=\{b\}$ and if $I_{a_0}\cap\Z=\{b_0\}$ then $\frac{b}{a}=\frac{b_0}{a_0}$.
\end{lemma}
\proof
By lemma \ref{the-2}, we have $\frac{f(\beta)}{\beta^{d_t}}-\frac{\varepsilon}{2}<\frac{b}{a}<\frac{f(\beta)}{\beta^{d_t}}+\frac{\varepsilon}{2}$, so
$\frac{f(\beta)}{\beta^{d_t}} a-\frac{\varepsilon}{2} a<b<\frac{f(\beta)}{\beta^{d_t}} a+\frac{\varepsilon}{2} a$, and the existence is proved. As the length of $(\frac{f(\beta)}{\beta^{d_t}} a-\frac{\varepsilon}{2} a,\frac{f(\beta)}{\beta^{d_t}} a+\frac{\varepsilon}{2} a)$ is $<2\frac{\varepsilon}{2} a\leq \varepsilon H\leq \frac{1}{H-1}\leq1$, so $b$ is the unique integer in the interval.

Assume that there is another $a_0\in\{1,2,\dots,H\}$, such that $(\frac{f(\beta)}{\beta^{d_t}} a_0-\frac{\varepsilon}{2} a_0,\frac{f(\beta)}{\beta^{d_t}} a_0+\frac{\varepsilon}{2} a_0)$ contains the integer $b_0$. Then $\frac{f(\beta)}{\beta^{d_t}} a_0-\frac{\varepsilon}{2} a_0<b_0<\frac{f(\beta)}{\beta^{d_t}} a_0+\frac{\varepsilon}{2} a_0$, so $\frac{f(\beta)}{\beta^{d_t}}-\frac{\varepsilon}{2}<\frac{b_0}{a_0}<\frac{f(\beta)}{\beta^{d_t}}+\frac{\varepsilon}{2}$. If $\frac{a}{b}\neq\frac{a_0}{b_0}$, then $|\frac{a}{b}-\frac{a_0}{b_0}|=|\frac{ab_0-a_0b}{bb_0}|\geq \frac{1}{H(H-1)}=\varepsilon$,
which contradicts to that the length of the interval is less than $\varepsilon$.\qed

Let $r_1:=\frac{f(\beta)}{\beta^{d_t}} -\frac{\varepsilon}{2},r_2:=\frac{f(\beta)}{\beta^{d_t}} +\frac{\varepsilon}{2}$.
By Lemma \ref{the-3}, if $a_0$ is the smallest positive integer such that $(a_0r_1,a_0r_2)$ contains the unique integer $b_0$, then we have $c_t=\frac{b_0}{a_0}$.
Note that $a_0$ is the smallest integer such that $(a_0r_1,a_0r_2)$ contains the unique integer $b_0$ if and only if
$a_0$ is the smallest integer such that  $b_0/a_0$ is in $(r_1,r_2)$,
and such an $a_0$ can be found with Algorithm \ref{alg-mindn}.
This observation leads to the following algorithm to find the leading coefficient of $f(x)$.
\begin{alg}[ULCoefRat]\label{alg-uvlcr}
\end{alg}

{\noindent\bf Input:} $f(\beta),\beta,\varepsilon,d_t$

{\noindent\bf Output:} the leading coefficient of $f(x)$.

\begin{description}
\item[Step 1] $\mathbf{if}$ $\frac{f(\beta)}{\beta^{d_t}}>0$, $\mathbf{then}$
$r_1:=\frac{f(\beta)}{\beta^{d_t}}-\frac{\varepsilon}{2}$,
$r_2:=\frac{f(\beta)}{\beta^{d_t}}+\frac{\varepsilon}{2}$;
$\mathbf{else}$
$r_1:=-\frac{f(\beta)}{\beta^{d_t}}-\frac{\varepsilon}{2}$,
$r_2:=-\frac{f(\beta)}{\beta^{d_t}}+\frac{\varepsilon}{2}$;

\item[Step 2:]Let $a:=$ {\bf MiniDenom} $(r_1,r_2)$;

\item[Step 3:] Return $\frac{\lceil a(\frac{f(\beta)}{\beta^{d_t}}-\frac{\varepsilon}{2})\rceil}{a}$

\end{description}

Replacing Algorithm {\bf{ULCoef}} with Algrothm {\bf{ULCoefRat}}
in Algorithm {\bf{UPolySI}}, we obtain the following interpolation algorithm for sparse polynomials with rational coefficients.
\begin{alg}[UPolySIRat]\label{alg-uprc}
\end{alg}
{\noindent\bf Input:} A black box polynomial $f(x)\in \Q[x]$ whose coefficients
are in $A$ given in \bref{eq-rat}.

{\noindent\bf Output:} The exact form of $f(x)$.
%
%
%\begin{description}
%\item[Step 1:]
%$\mathbf{If}$ $H>1$ $\mathbf{then}$ $\varepsilon:=\frac{1}{H(H-1)}$,
%%
%%$\mathbf{else}$ $H=1$ $\mathbf{then}$ $\varepsilon:=1$;
%$\mathbf{else}$ $\varepsilon:=1$;
%
%\item[Step 2:]
%Let $p:=\frac{2C}{\varepsilon}+1$.
%
%\item[Step 3:]
%Evaluate $f(p)$.
%
%\item[Step 4:]
%Let $g:=0,u:=f(p)$.
%
%\item[Step 5:]
%
%$\mathbf{while}$ $u\neq 0$ $\mathbf{do}$
%
% $d:=${\bf UVDeg}$(u,\varepsilon,p)$;
%
% $c:=${\bf LeadCoef}$(\frac{u}{p^d},\varepsilon)$;
%
%  $u:=u-cp^d$;
%
%   $g:=g+cx^d$;
%
%  $\mathbf{end\ do}$.
%
%\item[Step 6:]
%$\mathbf{Return}$ $g$.
%\end{description}

\begin{theorem}\label{the-11}
The arithmetic operations of Algorithm \ref{alg-uprc} are $\mathcal{O}(t\log H)$ and the bit complexity is $\mathcal{O}(td\log H(\log C+\log H))$, where $d$ is the degree of $f(x)$.
\end{theorem}
\proof
In order to obtain the degree, we need one log arithmetic operation in field $\Q$, while in order to obtain the coefficient, we need $\mathcal{O}(\log H)$ arithmetic operations, so the total complexity is $\mathcal{O}(t\log H)$.

Assume $f(\beta)=\frac{a_1}{h_1}\beta^{d_1}+\frac{a_2}{h_2}\beta^{d_2}+\cdots+\frac{a_t}{h_t}\beta^{d_t}$ and let $H_i:=h_1\cdots h_{i-1}h_{i+1}\cdots h_t$. Then we have

$$f(\beta)=\frac{a_1H_1\beta^{d_1}+a_2H_2\beta^{d_2}+\cdots+a_tH_t\beta^{d_t}}{h_1h_2\cdots h_t}$$
Then $|a_1H_1\beta^{d_1}+a_2H_2\beta^{d_2}+\cdots+a_tH_t\beta^{d_t}|\leq H^{t-1}C(\beta^{d_t}+\cdots+\beta+1)=H^{t-1}\frac{C}{\beta-1}(\beta^{d_t+1}-1)$,
so its bit length is $\mathcal{O}(t\log H+d\log C+d\log H)$.
It is easy to see that the bit length of $h_1h_2\cdots h_t$ is $\mathcal{O}(t\log H)$.
So the total bit complexity is $\mathcal{O}((t\log H)(t\log H+D\log C+D\log H))$. As $t\leq d$, the bit complexity  is $\mathcal{O}(td\log H(\log C+\log H))$.
\qed

\begin{cor}\label{cor-11}
If the coefficients of $f(x)$ are integers in $[-C,C]$, then
Algorithm \ref{alg-uprc} computes $f(x)$ with  arithmetic complexity $\mathcal{O}(t)$ and with bit complexity $\mathcal{O}(td\log C)$.
\end{cor}

\begin{remark}\label{rem-1}

The bit complexity of Algorithm \ref{alg-uprc} is $\widetilde{\mathcal{O}}(td)$,
which seems to be the optimal bit complexity for deterministic and exact interpolation algorithms
for a black box polynomial $f(x)\in Q[x]$.
For a $t$-sparse polynomial, $t$ terms are needed and the arithmetic complexity is at least $\mathcal{O}(t)$.
For $\beta\in \C$, we have $|f(\beta)|\leq C \frac{\beta^{d+1}-1}{\beta-1}$,
where $C$ is defined in (\ref{eq-e}). If $|\beta|\ne1$, then the height of $f(\beta)$ is $d|\log\beta|+\log C$ or $\widetilde{\mathcal{O}}(d)$.
For a deterministic and exact algorithm,  $\beta$ satisfying $|\beta|=1$ seems not usable.
So the bit complexity is at least $\widetilde{\mathcal{O}}(td)$.
For instance,
the height of the data in Ben-or and Tiwari's algorithm is already $\widetilde{\mathcal{O}}(td)$~\cite{1,813}.

\end{remark}

\section{Multivariate polynomial sparse interpolation with modified Kronecker substitution}

In this section, we give a deterministic and a probabilistic polynomial-time reduction of
multivariate  polynomial interpolation to univariate polynomial interpolation.
%
%
%The main ideal is that we use Kronecker's substitution to reduce the multivariate case into the univariate case, but we use a prime to reduce the degree.

\subsection{Find a good prime}
We will show how to find a prime number which can be used in the reduction.

We assume $f(x_1,x_2,\dots,x_n)$ is a multivariate polynomial in $\Q[x_1,x_2,\dots,x_n]$ with a degree bound $D$, a term bound $T$, and  $p$ is a prime.
We use the substitution
\begin{equation}\label{eq-sub1}
x_i=x^{\mathbf{mod}((D+1)^{i-1},p)},i=1,2,\dots,n.
\end{equation}
For convenience of description, we denote
\begin{equation}\label{eq-sub11}
f_{x,p}:=f(x,x^{\mathbf{mod}((D+1),p)},\dots,x^{\mathbf{mod}((D+1)^{n-1},p)}).
\end{equation}
Then the degree of $f_{x,p}$ is no more than $D(p-1)$ and the number of terms of $f_{x,p}$ is no more than $T$.

If the number of terms of $f_{x,p}$ is the same as that of $f(x_1,x_2,\dots,x_n)$,
there is no collision in different monomials and we call such prime as a {\em good prime} for $f(x_1,x_2,\dots,x_n)$.

If  $p$ is a good prime, then we can consider a new substitution:
\begin{equation}\label{eq-sub2}
x_i=q_ix^{\mathbf{mod}((D+1)^{i-1},p)},i=1,2,\dots,n,
\end{equation}
where $q_i,i=1,2,\dots,n$ is the $i$-th prime. In this case, each coefficient will change according to monomials of $f$.
%This method is proposed by Klivans and Spielman\cite{5}.
%
Note that in \cite{21}, the substitution is
$f(x,x^{(D+1)},\ldots,x^{(D+1)^{n-1}}) \mathbf{mod}(x^p-1)$.

%This method is proposed by Klivans and Spielman\cite{5}.
%
We show how to find a good prime $p$.
We first give a lemma.
\begin{lemma}\label{lm-4}
Suppose $p$ is a prime. If $\mathbf{mod}(a_1+a_2(D+1)+\cdots+a_n(D+1)^{n-1},p)\neq 0$, then $a_1+a_2\mathbf{mod}(D+1,p)+\cdots+a_n\mathbf{mod}((D+1)^{n-1},p)\neq 0$.
\end{lemma}
\proof
If $a_1+a_2\mathbf{mod}(D+1,p)+\cdots+a_n\mathbf{mod}((D+1)^{n-1},p)=0$, then $\mathbf{mod}(a_1+a_2(D+1)+\cdots+a_n(D+1)^{n-1},p)=0$, which contradicts to the assumption.
\qed

Now, we have the following theorem to find the good prime.

\begin{theorem}\label{th-gp1}
Let $f(x_1,x_2,\dots,x_n)$ be polynomial with  degree at most $D$ and $t\leq T$ terms. If
$$
N>\frac{T(T-1)}{2}\log_2[(D+1)^n-1]-\frac14 T^2+\frac12 T
$$
then there at least one of $N$ distinct odd primes $p_1,p_2,\dots,p_N$ is a good prime for $f$.
\end{theorem}
\proof
Assume $m_1,m_2,\dots,m_t$ are all the monomials in $f$, and $m_i=x_1^{e_{i,1}}x_2^{e_{i,2}}\cdots x_n^{e_{i,n}}$.
In order for $p$ to be a good prime, we need $e_{i,1}+e_{i,2}(\mathbf{mod}(D+1,p))+\cdots+e_{i,n}(\mathbf{mod}((D+1)^{n-1},p))\neq e_{j,1}+e_{j,2}(\mathbf{mod}(D+1,p))+\cdots+e_{j,n}(\mathbf{mod}((D+1)^{n-1},p))$, for all $i\neq j$. This can be change into $(e_{i,1}-e_{j,1})+(e_{i,2}-e_{j,2})(\mathbf{mod}(D+1, p))+\cdots+(e_{i,n}-e_{j,n})(\mathbf{mod}((D+1)^{n-1}, p))\neq 0$.
By Lemma \ref{lm-4}, it is enough to show $$\mathbf{mod}((e_{i,1}-e_{j,1})+(e_{i,2}-e_{j,2})(D+1)+\cdots+(e_{i,n}-e_{j,n})(D+1)^{n-1}, p)\neq 0, i\neq j$$
Firstly, $|(e_{i,1}-e_{j,1})+(e_{i,2}-e_{j,2})(D+1)+\cdots+(e_{i,n}-e_{j,n})(D+1)^{n-1}|\leq D(1+(D+1)+\cdots+(D+1)^{n-1})=(D+1)^n-1$.

We assume that $\overline{f}(x)=a_1x^{k_1}+a_2x^{k_2}+\cdots+a_tx^{k_t}$ is the polynomial after the Kronecker substitution, where $k_i=e_{i,1}+e_{i,2}(D+1)+\cdots+e_{i,n}(D+1)^{n-1}$.
If $t=2$, it is trivial. So now we assume $t>2$ and  we analyse how many kinds of primes the number $\prod_{i>j}(k_i-k_j)$ has. Without lose of generality, assume $k_1,k_2\dots,k_w$ are even, $k_{w+1},k_{w+2}\dots,k_t$ are odd, denote $v:=t-w$.  It is easy to see that $k_i-k_j$ has factor $2$ if $1\leq i\neq j\leq w$ or $w+1\leq i \neq j\leq t$.

If one of the $w$ and $v$ is zero, then $\prod_{i>j}(k_i-k_j)$ has a factor $2^{\frac{t(t-1)}{2}}$.

If both $w,v$ are not zero, then $\prod_{i>j}(k_i-k_j)$ has a factor $2^{\frac{w(w-1)}{2}+\frac{v(v-1)}{2}}$.

We give a lower bound of $\frac{w(w-1)}{2}+\frac{v(v-1)}{2}$.

As $\frac{w(w-1)}{2}+\frac{v(v-1)}{2}=\frac{w^2+v^2-t}{2}\geq\frac{1/2 (w+v)^2-t}{2}=\frac{1}{4}t^2-\frac{1}{2}t$, $\prod_{i>j}(k_i-k_j)$ at least has a factor $2^{\frac{1}{4}t^2-\frac{1}{2}t}$.

Since $|k_i-k_j|\leq (D+1)^n-1$, we have $\prod_{i>j}(k_i-k_j)\leq [(D+1)^n-1]^{\frac{t(t-1)}{2}}$.

If $p_1,p_2,\dots,p_N$ are distinct primes satisfying
 $$p_1p_2\dots p_N>\frac{[(D+1)^n-1]^{\frac{t(t-1)}{2}}}{2^{\frac{1}{4}t^2-\frac{1}{2}t}}$$
Then at least one of the primes is a good prime.
Since $p_i\geq 2$, $N>\frac{t(t-1)}{2}\log_2[(D+1)^n-1]-\frac14 t^2+\frac12 t$.

As we just know the upper bound $T$ of $t$, we can choose $T-t$ different positive integer $k_{t+1},k_{t+2},\dots,k_{T}$ which are different from $k_1,k_2,\dots,k_t$. So we still can use $T$ as the number of the terms. We have proved the lemma.\qed

\subsection{A deterministic algorithm}

\begin{lemma}\label{lm-203}
Assume $f=\frac{c_1}{H_1}x^{d_1}+\frac{c_2}{H_2}x^{d_2}+\dots+\frac{c_t}{H_t}x^{d_t}$, where $c_1,c_2,\dots,c_t\in\Z,H_1,H_2,\dots,H_t\in \Z_{+},d_1,d_2,\dots,d_t\in \N,d_1<d_2<\cdots<d_t,|\frac{c_i}{H_i}|\leq C$, $H_1,H_2,\dots,H_t,d_1,d_2,\dots,d_t$ are known. Let $H_{\max}:=\max\{H_1,H_2,\dots,H_t\}$. If $\beta\geq 2CH_{\max}+1$, then we can recover $c_1,c_2,\dots,c_t$ from $f(\beta)$.
\end{lemma}
\proof
It suffices to show that $c_t$ can be recovered from $f(\beta)$. As $\beta-1\geq 2CH_{\max}\geq 2CH_{t}$, then $\frac12\geq \frac{CH_t}{\beta-1}$. So $|f(\beta)H_t-c_t\beta^{d_t}|=|\frac{c_1H_t}{H_1}\beta^{d_1}+\frac{c_2H_t}{H_2}\beta^{d_2}+\dots+\frac{c_{t-1}H_t}{H_{t-1}}p^{d_{t-1}}|\leq CH_t (\frac{\beta^{d_t}-1}{\beta-1})\leq\frac12 (\beta^{d_t}-1)$.
So $|\frac{f(\beta)H_t}{\beta^{d_t}}-c_t|<\frac12$. That is $\frac{f(\beta)H_t}{\beta^{d_t}}-\frac12<c_t<\frac{f(\beta)H_t}{\beta^{d_t}}+\frac12$.
Since $c_t$ is an integer, $c_t=\lceil\frac{f(\beta)H_t}{\beta^{d_t}}-\frac12 \rceil$. The rest can be proved by induction.  \qed

\begin{alg}[MPolySIMK]\label{alg-mp1}
\end{alg}
{\noindent\bf Input:} A black box polynomial $f(x_1,x_2,\dots,x_n)\in A[x_1,x_2,\dots,x_n]$, whose coefficients
are in $A$ given in \bref{eq-rat}, an upper bound $D$ for the degree, an upper bound $T$ of the number of terms, a list of $n$ different primes $q_1,q_2,\dots,q_n(q_1<\cdots<q_n)$.

{\noindent\bf Output:} The exact form of $f(x_1,x_2,\dots,x_n)$.

\begin{description}

\item[Step 1:]
Randomly choose $N$ different odd primes $p_1,p_2,\dots,p_N$, where

 $N=\lfloor\frac{T(T-1)}{2}\log_2[(D+1)^n-1]-\frac14 T^2+\frac12 T\rfloor+1$.

\item[Step 2:]

$\mathbf{for}$ $i=1,2,\dots,N$ $\mathbf{do}$

Let $f_i:=\mathbf{UPolySIRat}(f_{x,p_i},A,T)$ via Algorithm \ref{alg-uprc}, where
$f_{x,p_i}$ is defined in \bref{eq-sub11}.

\item[Step 3:]
Let $S:=\{\}$;

$\mathbf{for}$ $i=1,2,\dots,N$ $\mathbf{do}$

if $f_i\neq failure$, then $S:=S\bigcup\{f_i\}$.

$\mathbf{end\ do}$;

\item[Step 4:]
$\mathbf{Repeat}$:

Choose one integer $i$ such that $f_{i}$ has the most number of the terms in $S$.

$\mathbf{if}$ $f_i(j)=f_{x,p_i}(j)$ for $j=1,2,\dots,D(p_i-1)+1$ $\mathbf{then}$ break $\mathbf{Repeat}$;

$S:=S\backslash \{f_i\}$

$\mathbf{end\ Repeat}$

Let $i_0$ be the integer found  and $f_{i_0}=\frac{c_1}{H_1}x^{d_1}+\frac{c_2}{H_2}x^{d_2}+\cdots+\frac{c_t}{H_t}x^{d_t},d_1<d_2<\cdots<d_t$

\item[Step 5:]

Let $\beta:=2Cq_n^D\max\{H_1,H_2,\dots,H_t\}+1$.[Lemma \ref{lm-203}]

Denote $g= f(q_1x,q_2x^{\mathbf{mod}(D+1,p_{i_0})},\dots,q_nx^{\mathbf{mod}((D+1)^{n-1},p_{i_0})})$.

Let $u:=g(\beta)$.

\item[Step 6:]
Let $h:=0$.

$\mathbf{for}$ $i=t,t-1,\dots,1$ $\mathbf{do}$

Let $b:=\lceil \frac{u}{\beta^{d_i}}H_i-\frac12\rceil$

Factor $\frac{b}{c_i}$ into $q_1^{e_1}q_2^{e_2}\cdots q_n^{e_n}$.

$h:=h+\frac{c_i}{H_i}x_1^{e_1}x_2^{e_2}\cdots x_n^{e_n}$.

$u:=u-\frac{b}{H_i}\beta^{d_i}$

$\mathbf{end\ do}$;

\item[Step 7:]
return $h$.
\end{description}

\begin{remark}\label{remar-1}
If $p_i$ is not a good prime for $f$, then the substitution $f_{x,p_i}$ of $f$ has collisions.
$f_{x,p_i}$ may have some coefficients not in $A$.
So
we need to modify Step 4 of Algorithm \ref{alg-uprc}  as follows, with $T$ as an extra input.
For $c=\frac ab$,
if $|c|>C$, $|b|>H$, or the number of the terms of $f_i$ are more than $T$, then we let $f_i=failure$.
\end{remark}

\begin{theorem}\label{th-algmp1}
Algorithm \ref{alg-mp1} is correct and its bit complexity is $\widetilde{\mathcal{O}}(n^2T^5D\log H\log C+n^2T^5D\log^2 H+n^3T^6D^2)$.
%, where $\widetilde{\mathcal{O}}$ to denote the complexity bound ignoring the logarithm factors
\end{theorem}
\proof
First, we show the correctness.
If $p_i$ is a good prime for $f $, then all the coefficients of $f_{x,p_i}$ are in $A$. So in step 2,  Algorithm \ref{alg-uprc} can be used to find $f_i=f_{x,p_i}$.
It is sufficient to show that the prime $p_{i_0}$ that corresponding to $i_0$ obtained in step $4$ is a good prime.
In step 4, if there exists a $j_0$ such that $f_i(j_0)\neq f_{x,p_i}(j_0)$, then $f_i\neq f_{x,p_i}$. This   only happens when some of the coefficients of $f_{x,p_i}$ are not in $A$.
That is, $p_i$ is not a good prime for $f $. So we throw it away.
If $f_{i_0}(j)=f_{x,p_{i_0}}(j)$ for $j=1,2,\dots,D(p_{i_0}-1)+1$ for some $i_0$.
Since $\deg f_{i_0}\leq D(p_{i_0}-1)$, we have $f_{i_0}=f_{x,p_{i_0}}$.

Assume by contradiction that  $p_{i_0}$ is not a good prime for $f$, then the number of terms of $f_{i_0}$ is less than that of $f$.
Since $S$ includes at least one $f_{i_1}$ such that $p_{i_1}$ is good prime for $f$, the number of terms in $f_{i_1}$ is more than $f_{i_0}$. It contradicts to that $f_{i_0}$ has the most number of the terms in $S$. So $p_{i_0}$ is a good prime for $f$.

As $f_{i_0}=\frac{c_1}{H_1}x^{d_1}+\frac{c_2}{H_2}x^{d_2}+\cdots+
\frac{c_t}{H_t}x^{d_t},d_1<d_2<\cdots<d_t$, we can assume $f=\frac{c_1}{H_1}m_1+\frac{c_2}{H_2}m_2+\cdots+
\frac{c_t}{H_t}m_t$, where $m_i=x_1^{e_{i,1}}x_2^{e_{i,2}}\cdots x_n^{e_{i,n}}$.
We can write $g$ as $g=f(q_1x,q_2x^{\mathbf{mod}(D+1,p_{i_0})},\\\dots,q_nx^{\mathbf{mod}((D+1)^{n-1},p_{i_0})})=\frac{c_1 q_1^{e_{1,1}}q_2^{e_{1,2}}\cdots q_n^{e_{1,n}}}{H_1}x^{d_1}+\frac{c_2 q_1^{e_{2,1}}q_2^{e_{2,2}}\cdots q_n^{e_{2,n}}}{H_2}x^{d_2}+\cdots+
\frac{c_t q_1^{e_{t,1}}q_2^{e_{t,2}}\cdots q_n^{e_{t,n}}}{H_t}x^{d_t}$.
Since $|\frac{c_i q_1^{e_{i,1}}q_2^{e_{i,2}}\cdots q_n^{e_{i,n}}}{H_i}|\leq Cq_n^D$, by Lemma \ref{lm-203}, in step 6, $b=c_i q_1^{e_{i,1}}q_2^{e_{i,2}}\cdots q_n^{e_{i,n}}$.
By factoring $\frac{b}{c_i}=q_1^{e_{i,1}}q_2^{e_{i,2}}\cdots q_n^{e_{i,n}}$, we obtain the degrees of $m_i$.
We have proved the correctness.

We now analyse the complexity.
In step 2, we call Algorithm $\mathbf{UPolySIRat}$ $\mathcal{O}(nT^2\log D)$ times. The degree of $f_{x,p_i}$ is bounded by $D(p_i-1)$. Since the $i$-th prime is $\mathcal{O}(i\log i)$ and we use at most $\mathcal{O}(nT^2\log D)$ primes,  the degree bound is $\widetilde{\mathcal{O}}(nT^2D)$. So by Theorem \ref{the-11}, the bit complexity of getting all $f_i$ is $\widetilde{\mathcal{O}}((nT^3D\log H)(\log C+\log H)(nT^2\log D))$, this is $\widetilde{\mathcal{O}}(n^2T^5D\log H\log C+n^2T^5D\log^2 H)$.

In step 4, since $\deg f_i$ is $\widetilde{\mathcal{O}}(nT^2D)$, by fast multipoint evaluation~\cite[p.299]{8}, it needs $\widetilde{\mathcal{O}}(nT^2D)$ operations.
The number of the $f_i$ that we need to check is at most $\widetilde{\mathcal{O}}(nT^2\log D)$, so the total arithmetic operation for evaluations is $\widetilde{\mathcal{O}}(n^2T^4 D)$. As the coefficients of $f_i$ are in $A$ and the number of terms  is less than $T$, the data is $\widetilde{\mathcal{O}}(TC(nT^2D)^{nT^2D}H^T)$. So the height of the data is $\widetilde{\mathcal{O}}(nT^2D+\log C+T\log H)$. The total bit complexity of step 4 is $\widetilde{\mathcal{O}}(n^3T^6D^2+n^2T^4D\log C+n^2T^5 D\log H)$.

In step 6, we need to obtain $t$ terms of $g$. We analyse the bit complexity of one step of the cycle. To obtain $b$, we need $\mathcal{O}(1)$ arithmetic operations. The height of the data is $\widetilde{\mathcal{O}}(nT^2D(\log C+D\log n+\log H))$, so the bit complexity is $\widetilde{\mathcal{O}}(nT^2D\log C+nT^2D^2+nT^2D\log H)$.
To factor $\frac{b}{c_i}$, we need $n\log^2 D$ operations. The data of $b$ and $c_i$ is $\widetilde{\mathcal{O}}(Cq_n^DH)$, so the bit complexity is $\widetilde{\mathcal{O}}(n\log^2 D\log C+nD+n\log^2D\log H)$.
So the total bit complexity of step 6 is $\widetilde{\mathcal{O}}(nT^3D\log C+nT^3D^2+nT^3D\log H)$.

Therefore, the bit complexity is $\widetilde{\mathcal{O}}(n^2T^5D\log H\log C+n^2T^5D\log^2 H+n^3T^6D^2)$.\qed

\begin{remark}\label{remar-10}
If $A=\{a|C\geq |a|,a\in \Z\}$, we can modified the Algorithm \ref{alg-mp1}. Assume $A_{T}=\{a|TC\geq |a|,a\in \Z\}$. In step 2, we let $f_i:=\mathbf{UPolySIRat}(f_{x,p_i},A_{T})$. As $f_{x,p_i}$ is an integer polynomial with coefficients bounded by $TC$, $f_i=f_{x,p_i}$. So in step 4, we just find the smallest integer $i_0$ that $f_{i_0}$ has the most number of the terms in $S$. In this case, $p_{i_0}$ is a good prime for $f$. The bit complexity of the algorithm will be $\widetilde{\mathcal{O}}(n^2T^5D\log C+nT^3D^2)$.
\end{remark}

\subsection{Probabilistic Algorithm}

Giesbrecht and Roche~\cite[Lemma 2.1]{6} proved that if $\lambda=\max\{21,\frac 5 3 nT(T-1)\ln D\}$,
then a prime $p$ chosen at random in   $[\lambda,2\lambda]$ is a good prime for $f(x_1, \dots,x_n)$ with probability at least $\frac1 2$.
Based on this result, we give a probabilistic algorithm.

\begin{alg}[ProMPolySIMK]\label{alg-pmp1}
\end{alg}
{\noindent\bf Input:} A black box polynomial $f(x_1, \dots,x_n)\in A[x_1, \dots,x_n]$, whose coefficients
are in $A$ given in \bref{eq-rat}, an upper bound $D$ for the degree, an upper bound $T$ of the number of terms, a list of $n$ different primes $q_1,q_2,\dots,q_n(q_1<\dots<q_n)$.

{\noindent\bf Output:} The exact form of $f(x_1, \dots,x_n)$ with probability $\ge\frac12$.

\begin{description}

\item[Step 1:]Let $\lambda:=\max\{21,\frac 5 3 nT(T-1)\ln D\}$,
randomly choose a prime $p$ in $[\lambda,2\lambda]$.

\item[Step 2:]
Let
$f_p:=\mathbf{UPolySIRat}(f_{x,p},A,T)$ via Algorithm \ref{alg-uprc}.

$\mathbf{if}$ $f_p=failure$ $\mathbf{then}$ return failure;

Assume $f_p=\frac{c_1}{H_1}x^{d_1}+\frac{c_2}{H_2}x^{d_2}+\cdots+\frac{c_t}{H_t}x^{d_t},d_1<d_2<\cdots<d_t$

\item[Step 3:]

Let $\beta:=2Cq_n^D\max\{H_1,H_2,\dots,H_t\}+1$.[Lemma \ref{lm-203}]

Denote $g(x)= f(q_1x,q_2x^{\mathbf{mod}(D+1,p)},\dots,q_nx^{\mathbf{mod}((D+1)^{n-1},p)})$.

Let $u:=g(\beta)$;

\item[Step 4:]
Let $s:=0$;

$\mathbf{for}$ $i=t,t-1,\dots,1$ $\mathbf{do}$

Let $b:=\lceil \frac{u}{\beta^{d_i}}H_i-\frac12\rceil$

Factor $\frac{b}{c_i}=kq_1^{e_1}q_2^{e_2}\cdots q_n^{e_n}$, where $q_i\nmid k,i=1,2,\dots,n$

$\mathbf{if}$ $k\neq 1$ or $e_1+e_2+\cdots+e_n>D$ $\mathbf{then}$ return failure;

$s:=s+\frac{c_i}{H_i}x_1^{e_1}x_2^{e_2}\cdots x_n^{e_n}$.

$u:=u-\frac{b}{H_i}x^{d_i}$

$\mathbf{end\ do}$;

$\mathbf{if}$ $u=0$ $\mathbf{then}$ return $s$

$\mathbf{else}$ return failure;

\end{description}

\begin{theorem}\label{th-qalgmp1}
The bit complexity of Algorithm \ref{alg-pmp1} is $\widetilde{\mathcal{O}}(nT^3D\log H\log C+nT^3D\log^2 H+nT^3D^2)$.
\end{theorem}
\proof
%In step 2, we need to call Algorithm \ref{alg-uprc}.
In step 2, the degree of $f_{x,p}$ is bounded by $D(p-1)$.  Since the $p$ is $\mathcal{O}(nT^2\log D)$, the degree bound is $\widetilde{\mathcal{O}}(nT^2D)$. By Theorem \ref{the-11}, the complexity is $\widetilde{\mathcal{O}}(( nT^3D\log H)(\log C+\log H))$, or $\widetilde{\mathcal{O}}(nT^3D\log H\log C+nT^3D\log^2 H)$.

In step 4, we need to obtain $t$ terms of $g$. We analyse the bit complexity of one step of the cycle. To obtain $b$, we need $\mathcal{O}(1)$ arithmetic operations. The height of the data is $\widetilde{\mathcal{O}}(nT^2D(\log C+D\log n+\log H))$, so the bit complexity is $\widetilde{\mathcal{O}}(nT^2D\log C+nT^2D^2+nT^2D\log H)$.
To factor $\frac{b}{c_i}$, we need $n\log^2 D$ operations. The height of $b$ and $c_i$ is $\widetilde{\mathcal{O}}(Cq_n^DH)$, so the bit complexity is $\widetilde{\mathcal{O}}(n\log^2 D\log C+nD+n\log^2D\log H)$.
So the total bit complexity of step 4 is $\widetilde{\mathcal{O}}(nT^3D\log C+nT^3D^2+nT^3D\log H)$.

Therefore, the total bit complexity of the algorithm  is $\widetilde{\mathcal{O}}(nT^3D\log H\log C+nT^3D^2+nT^3D\log^2 H)$.\qed

\begin{remark}
In Algorithm \ref{alg-pmp1}, we also modify   step 4 of Algorithm \ref{alg-uprc} as remark \ref{remar-1}.
\end{remark}

\section{Experimental results}

In this section, practical performances of the algorithms will be presented.
The data are collected on a desktop with Windows system,
3.60GHz Core $i7-4790$ CPU, and 8GB RAM memory.
The implementations in Maple can be found in
\begin{verbatim}
http://www.mmrc.iss.ac.cn/~xgao/software/sicoeff.zip
\end{verbatim}

%http://www.mmrc.iss.ac.cn/\~\,xgao/software/sicoeff.zip.

%In this subsection, we present the average running times for polynomials.
We randomly construct five polynomials, then regard them as black box polynomials
and reconstruct them with the algorithms. The average times are collected.

\begin{figure}[h]
\begin{minipage}[t]{0.49\linewidth}
\centering
\includegraphics[scale=0.22]{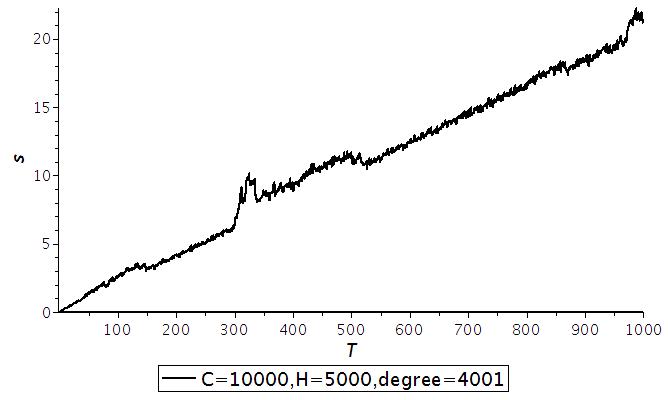}
\caption{$\mathbf{UPolySIRat}$: average running times with varying $T$} \label{fig1}
\end{minipage}\quad
\begin{minipage}[t]{0.48\linewidth}
\centering
\includegraphics[scale=0.21]{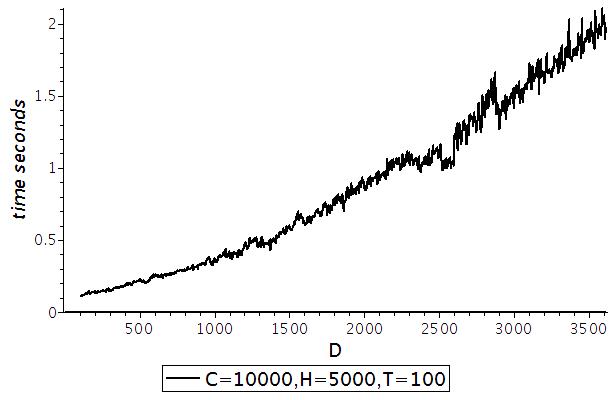}
\caption{$\mathbf{UPolySIRat}$: average running times with varying $D$ }\label{fig2}
\end{minipage}
\end{figure}

\begin{figure}[!hptb]
\begin{minipage}[t]{0.49\linewidth}
\centering
\includegraphics[scale=0.22]{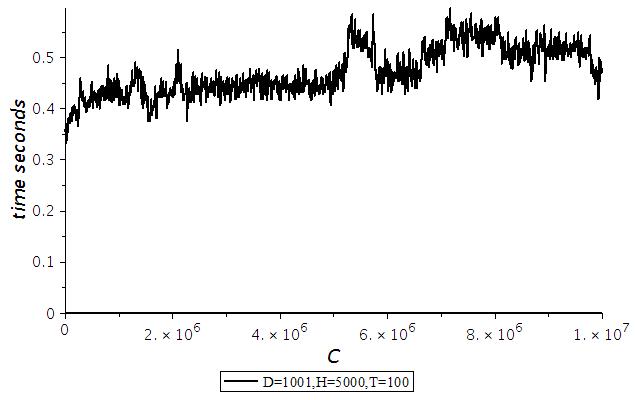}
\caption{$\mathbf{UPolySIRat}$: average running times with varying $C$} \label{fig3}
\end{minipage}\quad
\begin{minipage}[t]{0.48\linewidth}
\centering
\includegraphics[scale=0.21]{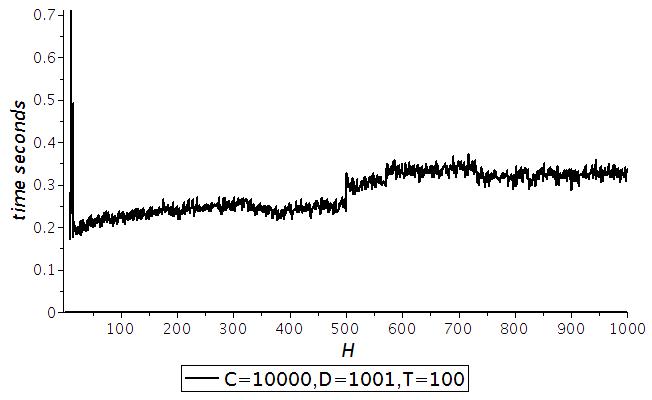}
\caption{$\mathbf{UPolySIRat}$: average running times with varying $H$ }\label{fig4}
\end{minipage}
\end{figure}

The results for univariate interpolation are shown in Figures \ref{fig1}, \ref{fig2}, \ref{fig3}, \ref{fig4}.
In each figure, three of the parameters $C,H,D,T$ are fixed and one of them is variant.
From these figures, we can see that  Algorithm $\mathbf{UPolySIRat}$  is linear in $T$, approximately linear in $D$, logarithmic in $C$ and $H$.

\begin{figure}[!hptb]
\begin{minipage}[t]{0.49\linewidth}
\centering
\includegraphics[scale=0.22]{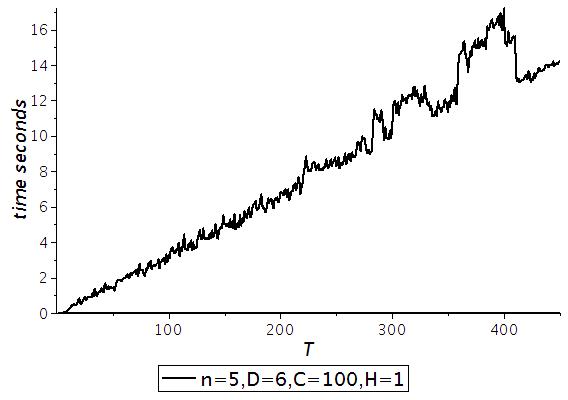}
\caption{$\mathbf{ProMPolySIMK}$: average running times with varying $T$} \label{fig5}
\end{minipage}\quad
\begin{minipage}[t]{0.48\linewidth}
\centering
\includegraphics[scale=0.21]{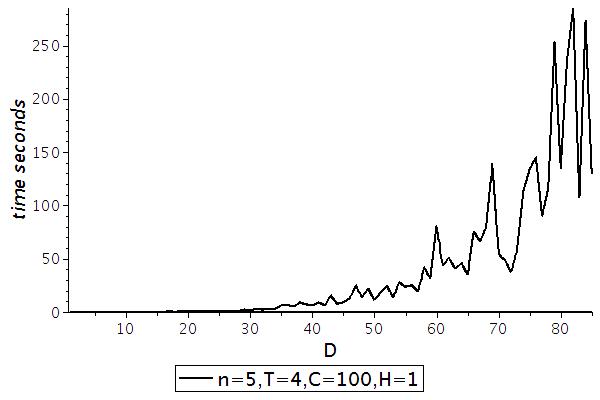}
\caption{$\mathbf{ProMPolySIMK}$: average running times with varying $D$ }\label{fig6}
\end{minipage}
\end{figure}

The results in the multivariate case are shown in Figures \ref{fig5}, \ref{fig6}. We just test the probabilistic algorithm.
From these figures, we can see that  Algorithm $\mathbf{ProMPolySIMK}$  are polynomial in $T$ and $D$.


\section{Conclusion}

In this paper, a new type of sparse interpolation is considered,
that is, the coefficients of the black box polynomial $f$
are from a finite set. Specifically, we assume that the
coefficients are rational numbers such that
the upper bounds of the absolute values of these numbers and
their denominators are given, respectively.
We first give an interpolation algorithm for a univariate polynomial $f$,
where $f$ is obtained from one  evaluation $f(\beta)$ for a sufficiently large number $\beta$.
Then, we introduce the modified Kronecker substitution
to reduce the interpolation of a multivariate polynomial into
the univariate case. Both algorithms have  polynomial bit-size complexity
and the algorithms can be used to recover quite large polynomials.


\begin{thebibliography}{99}

\bibitem{72}
A. Arnold.
Sparse Polynomial Interpolation and Testing.
PhD Thesis, Waterloo Unversity, Canada,2 016.

\bibitem{7}
A. Arnold and D.S. Roche.
Multivariate sparse interpolation using randomized Kronecker substitutions.
ISSAC'14, July 23-25, 2014, Kobe, Japan.

\bibitem{1}
M. Ben-Or and P. Tiwari.
A deterministic algorithm for sparse multivariate polynomial interpolation.
20th Annual ACM Symp. Theory Comp., 301-309, 1988.

\bibitem{21}
S. Garg and E. Schost.
Interpolation of polynomials given by straight-line programs.
Theoretical Computer Science, 410(27-29):2659-2662, 2009.

\bibitem{6}
M. Giesbrecht and D.S. Roche.
Diversification improves interpolation.
Proc. ISSAC'11, 123-130, ACM Press, 2011.

\bibitem{812}
Q.L. Huang and X.S Gao.
Sparse sational function interpolation with finitely many values for the coefficients.
arXiv:1706.00914, 2017.

\bibitem{8123}
Q.L. Huang and X.S Gao.
New algorithms for sparse interpolation and identity testing of multivariate polynomials.
Preprint, 2017.

\bibitem{813}
E. Kaltofen and L. Yagati.
Improved sparse multivariate polynomial interpolation algorithms.
Proc. ISSAC'88, 467-474, 1988.

\bibitem{5}
A.R. Klivans and D. Spielman.
Randomness efficient identity testing of multivariate polynomials.
In Proc. STOC '01, 216-223, ACM Press, 2001.

\bibitem{2}
L. Kronecker.
Grundz$\ddot{u}$ge einer arithmetischen theorie der algebraischen gr$\ddot{o}$ssen.
Journal f$\ddot{u}$r die reine und angewandte Mathematik, 92:1-122, 1882.

\bibitem{ls1}
Y.N. Lakshman and B.D. Saunders.
Sparse polynomial interpolation in nonstandard bases.
SIAM J. Comput., 24(2), 387-397, 1995.

\bibitem{8}
J. von zur Gathen and J. Gerhard.
Modern Computer Algebra.
Cambridge University Press, 1999.


\bibitem{71}
R. Zippel.
Interpolating polynomials from their values.
Journal of Symbolic Computation, 9(3), 375-403, 1990.
\end{thebibliography}
\end{document}